\documentclass[prl,twocolumn,superscriptaddress,showpacs]{revtex4}
\usepackage{epsfig}
\usepackage{amsmath}
\usepackage{amsfonts}

\begin{document}

\title{Fractional charge excitations in fermionic ladders.}

\author{B.N. Narozhny}
\affiliation{The Abdus Salam ICTP, Strada Costiera 11, 
Trieste, 34100, Italy.}

\author{S.T. Carr}
\affiliation{The Abdus Salam ICTP, Strada Costiera 11, 
Trieste, 34100, Italy.}

\author{A.A. Nersesyan}
\affiliation{The Abdus Salam ICTP, Strada Costiera 11, 
Trieste, 34100, Italy.}
\affiliation{The Andronikashvili Institute of Physics, Tamarashvili 6,
0177, Tbilisi, Georgia}

\date{\today}

\begin{abstract}
  
  The system of interacting spinless fermions hopping on a two-leg
  ladder in the presence of an external magnetic field is shown to
  possess a long range order: the bond density wave or the staggered
  flux phase. In both cases the elementary excitations are $Z_2$ kinks
  and carry one half the charge of an electron.

\end{abstract}


\maketitle

It has long been established both theoretically \cite{th1,bra} and
experimentally \cite{ex1} that quantum numbers of elementary
excitations in interacting many-body systems are not necessarily
limited to the values characterizing free particles. While excitations
carrying, for example, an electric charge double the free electron
charge $e$ can be visualized as pairs of the original particles
\cite{bcs}, the fractionally charged excitations may appear
counterintuitive \cite{qhe,bra,fab}. By far the most celebrated
example of such excitations are the fractionally charged
quasiparticles in the quantum Hall state \cite{qhe}. At about the same
time the notion of fractional charge appeared in the Peierls model
applied to quasi-one-dimensional conducting polymers \cite{bra}. More
recently \cite{fab}, fractionally charged excitations were predicted
in the context of the extended Hubbard model.

In this paper we argue that charge fractionalization is a generic
phenomenon in the sense that it appears already in the simplest model
of spinless fermions in a magnetic field. Restricting the fermions to
hop on a two-leg ladder\cite{cha,mik}, we show that just
nearest-neighbor interaction (if it is strong enough) leads to long
range order (LRO) with a doubly degenerate ground state. The elementary
excitations are then quantum domain walls or $Z_2$ kinks that carry
the charge $e/2$.

The applied field plays a crucial and somewhat surprising role in the
problem. In the absence of the field the only possible LRO is the
bond density wave (BDW) similar to that of Ref.~\onlinecite{fab}.
However in the presence of the field different types of LRO are
possible. The most spectacular manifestation of the field is the {\it
staggered flux phase} \cite{sfp,oaf} illustrated in
Fig.~\ref{result}. Note that the staggered order parameter here is
induced by a {\it uniform} field \cite{thi}.

{
\begin{figure}[ht]
\vspace{0.1 cm}
\epsfxsize=8 cm
\centerline{\epsfbox{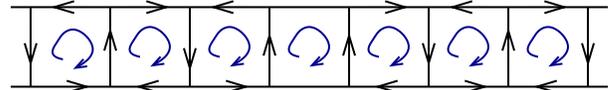}}
\vspace{0.1cm}
\caption{The staggered flux phase in the two-leg ladder penetrated by
a magnetic field. The arrows inside the plaquettes indicate the
direction of the applied uniform field (directed inside the plane of
the Figure). The arrows along the links indicate directions of the
staggered currents. The central plaquette is the domain wall between
the two degenerate ground states. This excitation carries the charge
$e/2$.}
\label{result}
\end{figure}
}

To build some intuition about the ordered phases consider the strong
coupling limit of the model where the transverse interaction $V_\perp$
is the strongest, so that no rung of the ladder can be doubly
occupied. For the quarter-filled ladder the starting point is a state
where there is one electron per two plaquettes. The hopping terms tend
to delocalize the electrons along the links. If the longitudinal
interaction is repulsive $V_{||}>0$, then the most favorable
configuration is that in which the electrons, avoiding to reside on
neighboring sites of the same chain, get delocalized on rungs thus
leading to a commensurate BDW. If, on the other hand, $V_{||}<0$ then
the system has a natural tendency to phase separate which is opposed
by the flux. If the flux is strong enough the electrons are placed on
every other plaquette and are completely delocalized around them (thus
avoiding double occupancy of the rungs) producing currents circulating
in the same direction (determined by the field). This can be viewed as
a staggered current superimposed upon the average, uniform
(persistent) current.

In this paper we show how the above cartoon picture arises starting
from the weak coupling limit. Our strategy is the following: (i) we
diagonalize the Hamiltonian in the absence of interaction; (ii)
linearize the exact spectrum near the Fermi points and bosonize the
model; (iii) finally, we solve the bosonic problem and find the ground
state of the interacting system.

The single-particle spectrum of the problem consists of two
one-dimensional bands and is governed by the magnetic flux through a
plaquette in units of the flux quantum $f$ as well as the ratio of the
transverse and longitudinal hopping amplitudes $\tau$. It is possible
to partially fill only one band and have only two Fermi points.  In
this case the low energy physics of the problem is similar to that of
the XXZ spin chain \cite{xxz}. If the interaction is weak, then the
system is in the Luttinger liquid regime (i.e. similar to the gapless
phase of the XXZ chain). If, on the other hand the two Fermi points
are commensurate with the lattice and the interaction is sufficiently
strong, then the umklapp scattering becomes relevant in the
renormalization group sense and opens a gap. The ground state of this
gapful phase possesses LRO and is doubly degenerate. Now, due to the
applied field, the physics of the ladder is richer than that of the
single chain. Depending on the sign of the umklapp term different
types of LRO are possible: BDW in the repulsive case, or the staggered
flux phase \cite{sfp} (or an orbital antiferromagnet \cite{oaf}) in
the case of attraction.

Having sketched our line of reasoning we now describe our
calculations. In this paper we focus on the particular parameter
regimes where the system exhibits the fractional charge. Mathematical
details and a description of the full phase diagram of the system as
well as a quantitative description of the strong coupling limit will
be discussed elsewhere \cite{us2}.

We start with the tight-binding Hamiltonian
\begin{eqnarray}
&&{\cal H} = 
-\sum\limits_n \Big[ \frac{1}{2}\sum\limits_{i=1,2}
\big( t_{||}(y_i)
c^\dagger_i(x_n) c_i(x_{n+1}) + h.c. \big)
\nonumber\\
&&
\nonumber\\
&& \quad\quad\quad
+ t_\perp c^\dagger_1(x_n) c_2(x_n)  + h.c.\Big] +{\cal H}_{int},
\label{h0} 
\end{eqnarray}
\noindent
where $c_j(x_n)$ is the electron annihilation operator on the chain
$j$ at the site $x_n$, $t_\perp$ and $t_{||}$ are the transverse and
the longitudinal hopping amplitudes. The magnetic field $B$ is
introduced by means of the Peierls substitution \cite{pei}.  Choosing
the Landau gauge \cite{dau} with the vector potential ${\bf A} = B(-y,
0, 0)$ and defining the $y$ coordinates of the chains as $y_{1(2)} =
\pm b/2$ we write the longitudinal hopping amplitude as
\begin{eqnarray}
t_{||}(y) = t_{||}^{(0)} e^{2\pi i f y/b}
\label{t}
\end{eqnarray}
\noindent
where $f = Bab/\phi_0$, $\phi_0$ being the flux quantum and $a$ the
lattice spacing along the chains. The control parameter $\tau$ is
defined as $\tau=t_\perp/t_{||}^{(0)}$; $t_{||}^{(0)}$ plays the role of
the bandwidth and hereafter will be set to unity. Expressed in terms
of the flux the model is explicitly gauge invariant.

{
\begin{figure}[ht]
%\vspace{0.1 cm}
\epsfxsize=5 cm
\centerline{\epsfbox{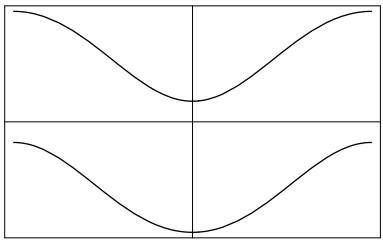}}
%\vspace{0.1cm}
\caption{The spectrum in the region $\tau > 1$, $\tau \cos\pi
  f>\sin^2\pi f $. }
\label{spectr1}
\end{figure}
}

The physical quantities we shall study in this paper are the
electrical current and particle density. In terms of the lattice
operators $c_j(x_n)$ the in-chain current is given
by
\begin{eqnarray}
J_{1(2)} = -\frac{i}{2} \left[ c_{1(2)}^\dagger(x_n) 
c_{1(2)}(x_{n+1}) e^{\mp\pi if}
- h.c. \right],
\label{j1}
\end{eqnarray}
\noindent
while the transverse current is defined as
\begin{eqnarray}
J_\perp = -i \tau\left[  c_1^\dagger(x_n) c_2(x_n) - h.c.
\right].
\label{jp}
\end{eqnarray}
\noindent
Similarly, the bond density on transverse links is 
\begin{eqnarray}
\rho_\perp = c_1^\dagger(x_n) c_2(x_n) + h.c.
\label{bd}
\end{eqnarray}
In the absence of interaction the Hamiltonian Eq.~(\ref{h0}) can be 
diagonalized with the help of the linear transformation
\begin{equation}
c_1(q) = u_q \alpha_q  + v_q \beta_q; \quad
c_2(q) = v_q \alpha_q  - u_q \beta_q;
\label{bt}
\end{equation}
\noindent
where the ``coherence factors'' are (the signs are explicit in
Eq.(\ref{bt}) so that the coherence factors are positive)
\begin{equation}
u_q^2(v_q^2) = \frac{1}{2}\left[1\mp\frac{\sin q \sin \pi f}
{\sqrt{\sin^2q\sin^2\pi f+\tau^2}}\right].
\label{cf}
\end{equation}
\noindent
The exact spectrum of the system consists of two one-dimensional bands
\begin{equation}
\epsilon_{\alpha(\beta)}(k)=-\cos k \cos\pi f \mp
\sqrt{\sin^2k\sin^2\pi f+\tau^2},
\label{spectrum}
\end{equation}
\noindent
where $k$ in the momentum along the chains. Since the transformation
$f\rightarrow 1-f$ preserves the spectrum we only need to consider the
fluxes such that $0\le f\le 1/2$.  If the transverse hopping amplitude
is large enough so that $\tau > \cos\pi f$ the spectrum possesses a
band gap and resembles a band insulator, see Fig.~\ref{spectr1}. If
the flux is not too small, $\sin^2\pi f > \tau\cos\pi f$, the bands
acquire a double-well shape shown in Fig.~\ref{spectr2}.

{
\begin{figure}[ht]
%\vspace{0.1 cm}
\epsfxsize=5 cm
\centerline{\epsfbox{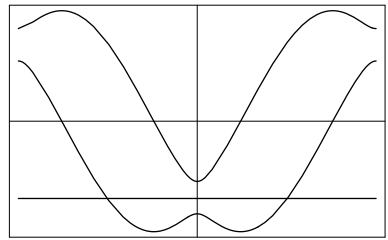}}
%\vspace{0.1cm}
\caption{The spectrum in the region 
  $(\sin\pi f)^2 /\tau > \cos\pi f > \tau$. The horizontal line depicts
  the chemical potential such that there are only two Fermi points.}
\label{spectr2}
\end{figure}
}

Now we focus on the situation where the lower band is only partially
filled (the upper band is empty) and there are only two Fermi
points. The ground state in the absence of interaction is
characterized by the persistent current flowing along the
chains. Indeed, the ground state expectation value of the transverse
current Eq.~(\ref{jp}) vanishes $\langle J_\perp\rangle = 0$, while
the in-chain current Eq.~(\ref{j1}) is non-zero as long as the flux is
applied
\begin{eqnarray}
\langle J_1\rangle =\frac{\sin\pi f}{2}
\sum\limits_{q=-k_F}^{k_F}\left[
\frac{\sin^2 q \cos\pi f}{\sqrt{\sin^2q\sin^2\pi f+\tau^2}}
-\cos q \right].
\label{pc}
\end{eqnarray}
\noindent
The current flows along the two chains in opposite directions $\langle
J_1\rangle =-\langle J_2\rangle$, so that the net current is equal to
zero as it should. The Fermi momentum in Eq.~(\ref{pc}) depends on the
applied field as
\begin{equation}
\cos k_F = -\tilde\mu \cos\pi f 
- \sqrt{\tau^2 + (1-\tilde\mu^2)\sin^2\pi f}, 
\label{kf}
\end{equation}
\noindent
where $\tilde\mu$ is the chemical potential in units of
$t_{||}^{(0)}$.

It is noteworthy that {\it all} (filled) states contribute to the
persistent current Eq.~(\ref{pc}). The persistent current in this
problem is not an infrared phenomenon and thus can not be accounted
for by an effective low energy theory. The low energy effects we
discuss below are thus taking place on the background of this {\it
single-particle} persistent current. We will not discuss further
corrections to Eq.~(\ref{pc}) in this paper.

To formulate an effective low energy theory we now linearize the exact
spectrum in the vicinity of the Fermi points Eq.~(\ref{kf}), assuming
that that the $k=0$ features of the exact spectrum are far away from
the Fermi level, the condition that has to be checked later. Since
only one band is filled, this step is fairly standard and is similar
to the case of a single chain\cite{xxz}. Separating fast and slow
variables, we introduce the left- and right-movers for the lower band
and find the standard kinetic term 
\begin{eqnarray}
{\cal H}_0 = v_F^\alpha
\sum\limits_k k \left(\alpha^\dagger_R(k)\alpha_R(k)
-\alpha^\dagger_L(k)\alpha_L(k)\right),
\label{ke}
\end{eqnarray}
\noindent
where the Fermi velocity is
\begin{equation}
v^\alpha_F = \sin k_F
\frac{\sqrt{(1-\tilde\mu^2)\sin^2\pi f + \tau^2}}
{\sqrt{\sin^2k_F\sin^2\pi f + \tau^2}}.
\label{vfa}
\end{equation}
\noindent
To obtain the final form of the effective theory we need to express
the physical observables in terms of the low energy fields (in doing
so we approximate the coherence factors by their values at the Fermi
points $u_0$ and $v_0$). The smooth part of the particle density is
then given by
\begin{eqnarray}
\rho_1=u_0^2 J_R + v_0^2 J_L; \quad
\rho_2=v_0^2 J_R + u_0^2 J_L,
\label{jsm}
\end{eqnarray}
\noindent
where $J_{R(L)}=:\alpha^\dagger_{R(L)}\alpha_{R(L)}:$.
Mostly we shall be interested in {\it staggered} operators
with expectation values vanishing in the absence of interaction. The
operators of interest are the staggered currents (defined on links)
%\begin{widetext}
\begin{subequations}
\begin{eqnarray}
&&j_{1(2)}(x_n, x_{n+1}) = 
\frac{(-1)^n}{2}u_0 v_0
\Big(\alpha^\dagger_L(x_n)\alpha_R(x_{n+1})e^{-i \pi f}   
\nonumber\\
&&
\nonumber\\
&& \quad\quad\quad
-
\alpha^\dagger_L(x_{n+1})\alpha_R(x_n)e^{i \pi f} - h.c.
\Big),
\label{j12le}
\end{eqnarray}
\begin{equation}
j_\perp(x_n)=it_\perp(-1)^n\left(v^2_0-u_0^2\right)
\left(\alpha^\dagger_L(x_n)\alpha_R(x_n)-h.c.
\right),
\label{jple}
\end{equation}
\label{jle}
\end{subequations}
%\end{widetext}
\noindent
and the staggered bond density
\begin{equation}
\rho^{(s)}_\perp (x_n)=(-1)^n
\left(\alpha^\dagger_L(x_n)\alpha_R(x_n)+h.c \right).
\label{sbd}
\end{equation}
\noindent
Current conservation requires $|\langle j_\perp\rangle | = 2|\langle
j_{1(2)}\rangle |$.

We now turn to the discussion of interaction effects. For simplicity
we consider the nearest-neighbor density-density interaction
\begin{eqnarray}
&&{\cal H}_{int} = \sum\limits_n \Big[ 
V_{||} \Big(n_1(x_n) n_1(x_{n+1})+ n_2(x_n) n_2(x_{n+1})\Big)
\nonumber\\
&&
\nonumber\\
&& \quad\quad\quad\quad\quad\quad\quad\quad\quad
+V_\perp n_1(x_n) n_2(x_n) \Big],
\label{hi}
\end{eqnarray}
\noindent
with $n_j = :c^\dagger_j c_j:$ (colons indicate normal
ordering). 

In terms of left- and right-movers (neglecting trivial
renormalizations of the Fermi velocity) we rewrite the interaction
Eq.~(\ref{hi}) in position space as
\begin{eqnarray}
\label{hirl}
&&{\cal H}_{int}\approx \sum\limits_i \Big[
g_1 :\alpha^\dagger_R(x_i)\alpha_R(x_i)
\alpha^\dagger_L(x_i)\alpha_L(x_i):
\\
&&
\nonumber\\
&& \quad
+ \Big(g_2 
:\alpha^\dagger_R(x_i)\alpha^\dagger_R(x_{i+1})
\alpha_L(x_i)\alpha_L(x_{i+1}):+ h.c. \Big)\Big];
\nonumber
\end{eqnarray}
\noindent
where the new interaction constants are \cite{f1}
\begin{subequations}
\begin{equation}
g_1=2V_{||}u^2_0v^2_0
\left(1-\cos 2k_F\right)+V_\perp(u^2_0-v^2_0)^2;
\label{g1}
\end{equation}
\begin{equation}
g_2=-2V_{||}e^{2ik_F}u^2_0v^2_0.
\label{g2}
\end{equation}
\label{g}
\end{subequations}
The first term in Eq.~(\ref{hirl}) is the density-density interaction
that can be taken into account exactly by means of bosonization. The
second term is the umklapp interaction that does not conserve momentum
and thus is only relevant when $k_F$ is commensurate with the lattice
(i.e. for the quarter-filled ladder).

In the commensurate case the Fermi momentum $k_F=\pi/2$ and the
chemical potential $\tilde\mu=-\sqrt{\sin^2\pi f +\tau^2}$. This is
consistent with the requirement of having only two Fermi points when
$2\tau >|\cos 2\pi f|/\cos\pi f$. In this case the
combinations of coherence factors in Eqs.~(\ref{hirl}) and (\ref{jle})
simplify to
\begin{equation}
u^2_0-v^2_0 = (\sin\pi f)/\tilde\mu \quad ; \quad\quad
u_0v_0=\tau/\tilde\mu.
\label{cfc}
\end{equation}
Now we use the standard bosonization procedure and express the
interaction Hamiltonian Eq.~(\ref{hirl}) in terms of a bosonic field
$\phi$ and the conjugate momentum $\Pi$. Combining it with the kinetic
term we arrive to the usual form of the Luttinger liquid Hamiltonian
(here $\alpha$ is the bosonic cut-off)
\begin{equation}
{\cal H} = \frac{v_F^\alpha}{2}
\left[K\Pi^2+\frac{1}{K}\left(\partial_x\phi\right)^2\right]
-\frac{g_2}{2\pi^2\alpha^2}\cos\sqrt{16\pi}\phi.
\label{hl}
\end{equation}
\noindent
where the Luttinger liquid parameter $K$ is determined by the
interaction constant Eq.~(\ref{g1}) as $K=[(1-g_1/2\pi
v_F^\alpha)/(1+g_1/2\pi v_F^\alpha)]^{1/2}$.

The behavior of the effective theory Eq.~(\ref{hl}) is well known
\cite{xxz}. At the critical value of the Luttinger liquid parameter
$K=1/2$ \cite{f3} the system exhibits a Berezinski-Kosterlitz-Thouless
(BKT) \cite{bkt} transition to a gapful phase with broken $Z_2$
symmetry and LRO. The type of ordering depends on the sign of the
coupling constant $g_2$:

(i) If the umklapp term is repulsive $g_2>0$ then the local minima of
the potential are achieved at $\phi=\sqrt{\pi/16}+n\sqrt{\pi/4}$,
$n=0, \pm 1, \dots$. In this case the staggered bond density
Eq.~(\ref{sbd}) $\rho^{(s)}_\perp \sim \sin\sqrt{4\pi}\phi$ acquires a
non-zero expectation value and LRO is of the BDW-type. This ordering
resembles the charge density wave in the extended Hubbard model
\cite{fab} at $1/4$ filling.

(ii) If, on the other hand, $g_2<0$ \cite{f2}, then the local minima
are at $\phi=n\sqrt{\pi/4}$ and now there exists the staggered current
Eq.~(\ref{jle}) $j_\perp\sim\cos\sqrt{4\pi}\phi$ leading to the
staggered flux phase illustrated in Fig.~\ref{result}. Clearly, the
currents are non-zero only in the presence of the applied field, as
$j_\perp\propto\sin\pi f$. This is a surprise: we find that the {\it
uniform} field causes the {\it staggered} current.

In both cases the ground state is doubly degenerate and the elementary
excitation is a ``quantum domain wall'' or, in terms of the effective
theory Eq.~(\ref{hl}), a $Z_2$ kink interpolating between the two
neighboring minima. The charge carried by the kink is related to the
distance $\Delta\phi$ between the minima [as follows from
Eq.~(\ref{jsm})]
\begin{equation}
Q=e\frac{\Delta\phi}{\sqrt{\pi}}.
\label{q}
\end{equation}
\noindent
Since both types of ordering originate from the same interaction term
in Eq.~(\ref{hl}), the distance between the minima is the same for
both cases, $\Delta\phi = \sqrt{\pi/4}$, which brings us to the
conclusion that the kinks carry fractional charge
\begin{equation}
Q=\pm\frac{e}{2}.
\label{res}
\end{equation}
\noindent
This statement is the main result of the present paper. Three features
of the model were necessary to obtain this result: (i) the electron
density is bosonized as $\rho =\partial_x\phi/\sqrt{\pi}$; (ii) there
were two Fermi points with $2k_F=\pi$; (iii) the interaction was
strong enough to open the umklapp gap and produce the LRO.

To summarize, we have considered the simplest possible model of
interacting electrons subject to an external magnetic field
restricting the motion of electrons to hopping on a two-leg
ladder. Despite its apparent simplicity, the model exhibits phase
transitions to non-trivial LRO phases, the BDW phase in the case
$V_{||}>0$ and the staggered flux phase for $V_{||}<0$ \cite{f2}. In
both cases (due to their topological nature) elementary excitations
(i.e. kinks) carry fractional charge. As such, charge
fractionalization appears to be a generic phenomenon mostly
insensitive to the details of the model. As long as the basic features
of the model (i.e. the transverse hopping and the flux) are in place
any interaction that contains umklapp processes \cite{f1} will lead to
charge fractionalization.

The applied magnetic field played a crucial role in our
considerations. It led to a non-trivial deformation of the
single-particle spectrum (Fig.~\ref{spectr2}) previously
unaccounted for and to the appearance of the staggered flux phase. The
Fermi momentum became strongly dependent on the flux. For a situation
close to half-filling this implies the existence of four essentially
different Fermi points. This and other features of the model that
were not discussed in the present paper, as well as further
generalizations of the model to larger number of chains and spinful
electrons is the subject of a future publication \cite{us2}.

We are grateful to N. Andrei, T. Giamarchi, and A.M. Tsvelik for
stimulating discussions.

\end{document}